\documentstyle[epsfig, sprocl]{article}

\newcommand{\VL}{\left( \begin{array}{c}}
\newcommand{\VR}{\end{array} \right)}
\newcommand{\ML}{\left( \begin{array}{cc}}
\newcommand{\MLv}{\left( \begin{array}{cccc}}
\newcommand{\MR}{\end{array} \right)}
\newcommand{\dd}{\partial}

\newcommand{\hc}{\mbox {h.c.}}
\newcommand{\re}{\mbox {Re}}

\newcommand{\mste}{m_{\tilde{t}_1}}
\newcommand{\mstz}{m_{\tilde{t}_2}}

\newcommand{\msbl}{m_{\tilde{b}_L}}
\newcommand{\msbe}{m_{\tilde{b}_1}}

\newcommand{\MstL}{M_{\tilde{t}_L}}
\newcommand{\MstR}{M_{\tilde{t}_R}}
\newcommand{\MsbL}{M_{\tilde{b}_L}}
\newcommand{\MsbR}{M_{\tilde{b}_R}}
\newcommand{\MsqL}{M_{\tilde{q}_L}}
\newcommand{\MsqR}{M_{\tilde{q}_R}}
\newcommand{\mtlr}{M_t^{LR}}
\newcommand{\mblr}{M_b^{LR}}
\newcommand{\mqlr}{M_q^{LR}}

\newcommand{\mh}{m_h}
\newcommand{\pe}{\Phi_1}
\newcommand{\pz}{\Phi_2}

\newcommand{\msfi}{m_{\tilde{f}_i}}

\newcommand{\msq}{m_{\tilde{q}}}
\newcommand{\msqi}{m_{\tilde{q}_i}}
\newcommand{\msqe}{m_{\tilde{q}_1}}
\newcommand{\msqz}{m_{\tilde{q}_2}}

\newcommand{\SU}{{\mathrm SUSY}}
\newcommand{\oas}{{\cal O}(\alpha_s)}

\newcommand{\mt}{m_{t}}

\newcommand{\mq}{m_{q}}

\newcommand{\mgl}{m_{\tilde{g}}}
\newcommand{\sq}{\tilde{q}}
\newcommand{\sql}{\tilde{q}_L}
\newcommand{\sqr}{\tilde{q}_R}
\newcommand{\sqe}{\tilde{q}_1}
\newcommand{\sqz}{\tilde{q}_2}
\newcommand{\Stop}{\tilde{t}}
\newcommand{\Sbot}{\tilde{b}}
\newcommand{\sfl}{\tilde{f}_L}
\newcommand{\sfr}{\tilde{f}_R}

\newcommand{\sfi}{\tilde{f}_i}
\newcommand{\tst}{\theta_{\tilde{t}}}
\newcommand{\tsb}{\theta_{\tilde{b}}}
\newcommand{\tsq}{\theta_{\tilde{q}}}
\newcommand{\sw}{s_W}
\newcommand{\cw}{c_W}

\newcommand{\KL}{\left(}
\newcommand{\KR}{\right)}

\newcommand{\Tb}{\tan \beta\hspace{1mm}}

\newcommand{\CTb}{\cot \beta\hspace{1mm}}

\newcommand{\Sa}{\sin \alpha\hspace{1mm}}

\newcommand{\Ca}{\cos \alpha\hspace{1mm}}

\newcommand{\BE}{\begin{equation}}
\newcommand{\EE}{\end{equation}}
\newcommand{\BEA}{\begin{eqnarray}}
\newcommand{\BEAnn}{\begin{eqnarray*}}
\newcommand{\EEA}{\end{eqnarray}}
\newcommand{\EEAnn}{\end{eqnarray*}}
\newcommand{\non}{\nonumber}
\def\al{\alpha}

\def\de{\delta}
\def\eps{\varepsilon}

\def\De{\Delta}

\begin{document}

\thispagestyle{empty}
\setcounter{page}{0}
\def\thefootnote{\fnsymbol{footnote}}

\begin{flushright}
KA--TP--18--1997\\
hep-ph/9803294
\date{\today}
\end{flushright}

\vspace{1cm}

\begin{center}

{\large\sc {\bf Two--loop calculations in the MSSM}}
\footnote{Talk given at the workshop ``Quantum Effects in the MSSM'',
Sept. 1997, Barcelona, Spain}

\vspace{1cm}

{\sc S.~Heinemeyer
}

\vspace*{0.5cm}

Institut f\"ur theoretische Physik, Universit\"at
Karlsruhe,\\ 76128 Karlsruhe, Germany\\E-mail:
Sven.Heinemeyer@physik.uni-karlsruhe.de

\end{center}

\vspace*{1cm}

\begin{abstract}

Recent results on two--loop calculations in the MSSM are reviewed.
The computation of the QCD corrections to $\De\rho$, $\De r$
and to the mass of the lightest Higgs boson in the MSSM are
presented. 
 
\end{abstract}

\def\thefootnote{\arabic{footnote}}
\setcounter{footnote}{0}

\newpage


\title{TWO--LOOP CALCULATIONS IN THE MSSM}

\author{ S. HEINEMEYER}

\address{Institut f\"ur theoretische Physik, Universit\"at
Karlsruhe,\\ 76128 Karlsruhe, Germany\\E-mail:
Sven.Heinemeyer@physik.uni-karlsruhe.de }


\maketitle\abstracts{
Recent results on two--loop calculations in the MSSM are reviewed.
The computation of the QCD corrections to $\De\rho$, $\De r$
and to the mass of the lightest Higgs boson in the MSSM are
presented. 
}


\section{Introduction}
\subsection{Motivation}\label{subsec:motivation}

Supersymmetry (SUSY), in particular the Minimal Supersymmetric Standard
Model (MSSM), \cite{mssm}
is widely considered as the theoretically most 
appealing extension of the Standard Model (SM). It predicts the
existence of scalar partners $\sfl, \sfr$ to each SM chiral fermion,
and spin-1/2 partners to the gauge bosons and to the scalar Higgs
bosons. So far, the direct search for SUSY particles 
has not been successful. One can only set lower bounds of
${\cal O} (100)$ GeV on their masses~\cite{pdg}. 

An alternative way to probe the MSSM is to search for virtual
effects of the additional particles by means of precision
observables. Here most of the one--loop calculations are already
available~\cite{deltaroneloop,zobservoneloop}. 
In order to reach the same standard of theoretical predictions as in
the SM,
two--loop calculations are required.

Another way to test the MSSM is to search for the lightest Higgs
boson, which is at tree--level predicted to be lighter than the
Z~boson. The one--loop corrections are known to be huge
\cite{higgsmassoneloop}.
Hence a two--loop calculation is inevitable to have the mass
predictions under control.

Since the dominant two--loop contributions for electroweak precision
observables as well as for the Higgs sector are expected to originate from
the QCD 
corrections, we concentrate on $\oas$ calculations in this article.


\subsection{Notations}\label{subsec:notations}

Contrary to the SM, in the MSSM two Higgs doublets are needed.
The  Higgs potential is given by~\cite{hhg}
\BEA
\label{Higgspot}
V =& &m_1^2 H_1\bar{H}_1 + m_2^2 H_2\bar{H}_2 - m_{12}^2 (\epsilon_{ab}
      H_1^aH_2^b + \hc) \\ \nonumber
   &+&\frac{g'^2 + g^2}{8}\, (H_1\bar{H}_1 - H_2\bar{H}_2)^2
      +\frac{g^2}{2}\, |H_1\bar{H}_2|^2,
\EEA
where $m_i, m_{12}$ are soft SUSY-breaking terms, and
$g, g'$  are the $SU(2)$ and $U(1)$ gauge couplings.

\noindent
The doublet fields $H_1$ and $H_2$ are decomposed  in the following way:
\BEA
H_1 &=& \VL H_1^1 \\ H_1^2 \VR = \VL v_1 + (\Phi_1^{*0} + i\chi_1^{*0})
                                 /\sqrt2 \\ \Phi_1^- \VR ,\non \\
H_2 &=& \VL H_2^1 \\ H_2^2 \VR =  \VL \Phi_2^+ \\ v_2 + (\Phi_2^0 
                                     + i\chi_2^0)/\sqrt2 \VR.
\EEA
The vacuum expectation values $v_1$ and $v_2$ define
the angle $\beta$ via
\BE
\Tb = \frac{v_2}{v_1}.
\EE
In order to obtain the CP-even neutral
 mass eigenstates,  the rotation 
\BEA
\VL H^0 \\ h^0 \VR &=& \ML \Ca & \Sa \\ -\Sa & \Ca \MR 
\VL \Phi_1^0 \\ \Phi_2^0 \VR ,\\ \nonumber
\EEA
is performed. 
The potential (\ref{Higgspot}) can be described with the help of  two  
independent parameters: $\Tb$ and $M_A^2 = -m_{12}^2(\Tb+\CTb)$,
where $M_A$ is
the mass of the CP--odd $A$ boson.

SUSY associates a left-- and a right--handed
scalar partner to each SM quark. The current eigenstates, $\sql$ and
$\sqr$, mix to give the mass eigenstates $\sqe$ and $\sqz$.
In the MSSM the $\sq$
mass eigenstates and their mixing angles  are determined by
diagonalizing the following mass matrix ($e_q$ and $I_3^q$ are the
electric charge and the weak isospin of the partner quark,
$\MsqL, \MsqR$ and $A_q$ are soft SUSY-breaking terms)
\begin{equation}
{\cal M}^2_{\sq} = 
\left( 
  \begin{array}{cc} \MsqL^2 + \mq^2 + \cos 2 \beta (I_3^q
                       - e_q\sw^2) \, M_Z^2  & \mq \, \mqlr \\
                    \mq \, \mqlr & \MsqR^2 + \mq^2
                                   + e_q \cos 2 \beta \; \sw^2 \, M_Z^2
  \end{array}
\right)
\label{eq:stopmassmatrix}
\end{equation}
with $\sw^2=1-\cw^2 \equiv \sin^2 \theta_W$. 
Especially for the $\Stop-\Sbot$ sector the parameters are given by
$\mtlr = A_t - \mu \, \CTb$ and $\mblr = A_b - \mu \Tb$.
Furthermore, SU(2) gauge invariance
requires $M_{\tilde{t}_L} = M_{\tilde{b}_L}$ at the tree--level.

Due to the large value of the top--quark mass $\mt$, the mixing between the 
left-- and right--handed top squarks $\Stop_L$ and $\Stop_R$ can be
very large.


\section{Evaluation of the two--loop diagrams}

For precision observables like the $\rho$ parameter and $\De r$, which
fixes the higher order relation between $M_W, M_Z, G_F$ and $\al$, the
leading contributions can be expressed in terms of self--energies. The
corrections to the Higgs masses are given partly in terms of
self--energies and partly in terms of tadpole diagrams. Hence for the
corrections presented in this paper, it is sufficient to concentrate
on one-- and two--point functions.

One problem in dealing with two--loop calculations is the large number
of topologies and the even much larger number of diagrams. Thus we
make extensive use of computer algebra programs:
The two--loop diagrams for a specific process, including also the
required counterterm diagrams, are generated with the Mathematica
package {\em FeynArts} \cite{FA}. The model file we programmed for
this purpose  contains, besides the SM
propagators and vertices, the relevant part of the MSSM Lagrangian,
i.e.\ all SUSY propagators ($\Stop_1, \Stop_2, \Sbot_1, \Sbot_2,
\tilde{g}$) needed for the QCD--corrections and the appropriate
vertices
(gauge boson--squark vertices, squark--gluon and squark--gluino vertices). 
{\em FeynArts} inserts propagators and vertices into the 
graphs in all possible ways and creates the amplitudes including all
symmetry factors.

As shown in~\cite{TC}, an arbitrary two--loop self--energy diagram
can be reduced to a set of scalar one-- and two--loop integrals
by means of two--loop tensor integral decomposition.
Besides the well known functions $A_0$ and $B_0$
\cite{PV}, this set consists of the functions $T_{12345}, T_{1234},
T_{234}, T_{134}$, defined as~\cite{TC}
\BE
T_{12\dots 5} (k^2, m_1, m_2, \dots, m_5) = \int d^Dq_1 d^Dq_2
   \frac{1}{[k_1^2 - m_1^2] [k_2^2 - m_2^2] \dots [k_5^2 - m_5^2]},
\EE
with $k_i$ as internal momenta, $m_i$ as internal masses and $q_{1,2}$
as integration momenta.
The decomposition of the two--loop diagrams and counterterms
is performed  with the Mathematica package {\em TwoCalc}~\cite{TC}.

The scalar integrals can be split into a divergent and a finite
part, performing an expansion in $\de = (4 - {\rm D})/2$ (where D is
the dimension of space--time). This allows, after including the proper
counterterm diagrams, an
algebraical check of the finiteness. The finite result can now be
transformed into a FORTRAN code, which makes a fast numerical
evaluation possible.


\section{Dimensional Reduction (DRED)}\label{sec:dred}

\subsection{What is the problem?}

In SUSY theories the number of bosonic
degrees of freedom is equal to the number of the fermionic ones. 
If the regularization is performed through dimensional regularization
(DREG), all the fields are treated D-dimensional. A problem arises
through the fact that a vector in D dimensions has a different number
of d.o.f. than a vector in 4 dimensions, thus DREG breaks SUSY.

As a solution DRED was proposed \cite{dred}. Here the indices of the
fields (and corresponding matrices) are kept 4-dimensional, the
integrals and momenta are treated D-dimensional. Alternatively, it is
possible to use the so called $\eps$--scalars \cite{epsscalar}. 

For the program {\em TwoCalc} we implemented the option to
perform the calculation either in DREG or in DRED.

\subsection{What is changing?}

As explained in the previous chapter, a diagram is decomposed into a
set of basic scalar integrals. For the coefficients of these integrals
the calculation in DRED can give an extra contribution $\sim\de$ 
with respect to DREG. The changes for a diagram are the following:

\begin{itemize}

\item{$\frac{1}{\de^2}$--divergencies:} no change from DREG to DRED.
\item{$\frac{1}{\de}$--divergencies:} for one--loop diagrams no change
occurs, for two--loop diagrams, however, there is an extra
contribution, i.e. it is possible that a calculation is divergent in
DREG but finite in DRED (see Sec. \ref{sec:higgsmass}).
\item{finite part:} for one--loop diagrams as well as for two--loop
diagrams there arises an extra contribution. Thus it is possible that a
one--loop calculation is finite in DREG and in DRED, but has different
results.

\end{itemize}


\section{Renormalization}\label{sub:renormalization}

For our two--loop calculations for precision observables,
renormalization is required up to the two--loop level.

\begin{itemize}

\item{Gauge boson sector:} Two--loop renormalization is required. For
the W--boson we have
\BEA
\mbox {mass renormalization:} && M_W^2 \to M_W^2 + \de M_{W,1}^2 
                                 + \de M_{W,2}^2 \non \\
\mbox {field renormalization:} && W^\pm \to Z_W^{1/2} W^\pm \\
\mbox {with } Z_W^{1/2} = (1 + \de Z_{W,1} &+& \de Z_{W,2})^{1/2} 
       = 1 + \frac{1}{2} \de Z_{W,1} - \frac{1}{8} \de Z_{W,1}^2
                                     + \frac{1}{2} \de Z_{W,2} \non
\EEA

In the on--shell renormalization scheme one has:
\BEA
\de M_{W,1}^2 &=& \re \Sigma_{W,1}(M_W^2), \non \\
\de M_{W,2}^2 &=& \re \Sigma_{W,2}(M_W^2) + \mbox{combination of
one--loop terms}, \non \\
\de Z_{W,1} &=& -\re \Sigma'_{W,1}(M_W^2), 
\quad (\Sigma'(k^2) := \frac{\dd}{\dd k^2}\Sigma(k^2))\\
\de Z_{W,2} &=& -\re \Sigma'_{W,2}(M_W^2) + \mbox{combination of
one--loop terms}. \non
\EEA
For the Z this is performed analogously.

\item{Quark sector:} One--loop renormalization as in the SM has to be
inserted. 

\item{Squark sector:} One--loop renormalization is sufficient.
It can be performed in terms of masses and
mixing angle:
\BE
\msqi^2 \to \msqi^2 + \de\msqi^2, \quad
\tsq \to \tsq + \de\tsq.
\EE
For the masses on--shell conditions are used. The
renormalized mixing angle can be defined by requiring that the renormalized
squark mixing self--energy $\Sigma_{\sqe \sqz}^{\rm ren}(q^2)$ vanishes
when the $\sqe$ is on-shell:
\BE
\de\msqi^2 = \Sigma_{\sfi}(\msfi^2), \quad
\de\tsq = \frac{1}{\msqe^2 - \msqz^2}\Sigma_{\sqe,\sqz}(\msqe^2).
\EE
However, one has to keep in mind that there are relations between the
soft SUSY--breaking terms in the squark mass matrices, which also have
to be renormalized for a two--loop calculation (see~\cite{delrholong}).

\end{itemize}


\section{QCD corrections to $\De\rho$}\label{sec:delrho}

\subsection{Definition and one--loop results}\label{subsec:delrho:one} 

The deviation of the $\rho$ parameter from unity parametrizes the
leading universal corrections induced by heavy fields in electroweak
amplitudes, caused by a mass splitting between the partners in an
isospin doublet. Compared to this correction all additional
contributions are suppressed. $\De\rho$ gives an
important contribution to electroweak observables via
\BE
\De(\De r) \approx -\frac{\cw^2}{\sw^2}\De\rho, \quad
\De M_W \approx \frac{M_W}{2}\frac{\cw^2}{\cw^2 - \sw^2}\De\rho, \non
\EE
\BE
\De \sin^2\vartheta_W^{\rm eff} \approx 
                     - \frac{\cw^2 \sw^2}{\cw^2 - \sw^2}\De\rho.
\label{eq:delrhoapprox}
\EE

In terms of the transverse parts of the W-- and Z--boson self--energies
at zero momentum--transfer, the $\rho$
parameter is given by
\BE
\rho = \frac{1}{1-\De\rho} \ ; \ \ \ \ \De\rho =
\frac{\Sigma_{Z}(0)}{M_Z^2} - \frac{\Sigma_{W}(0)}{M_W^2}  \ .
\EE

\begin{figure}[htb]
\begin{center}
\mbox{
\psfig{figure=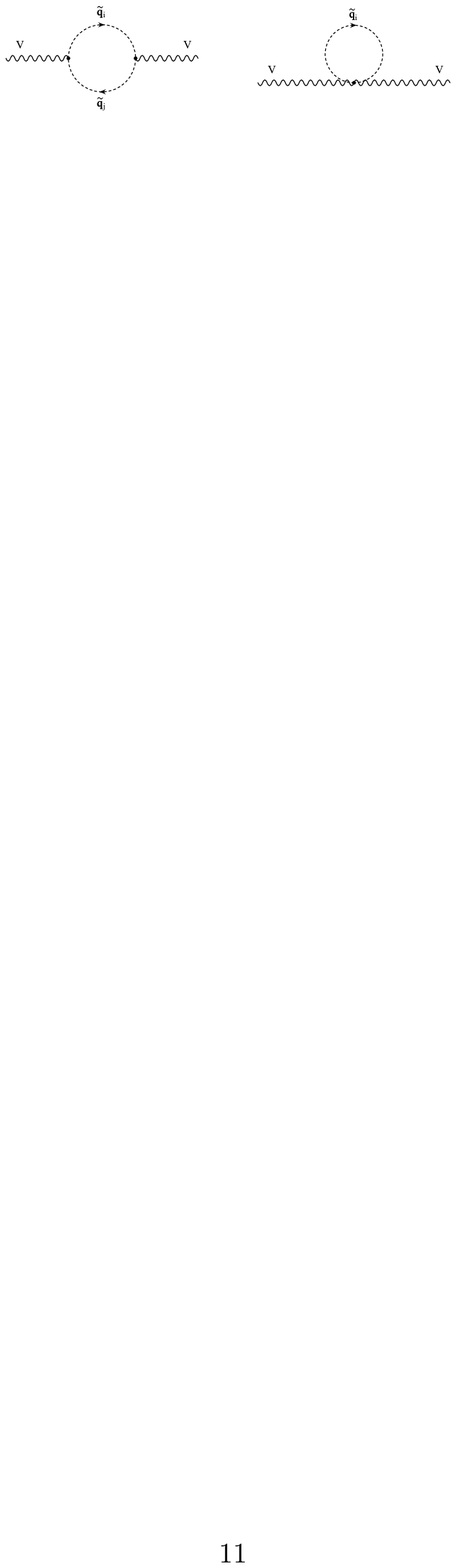,width=9cm,height=2cm,
                      bbllx=210pt,bblly=680pt,bburx=397pt,bbury=720pt}}
\end{center}
\caption[]{Feynman diagrams for the contribution of scalar quark loops 
to the gauge boson self--energies at one--loop order.}
\label{fig:oneloopdiagrams}
\end{figure}

Inserting the MSSM contribution to the W-- and Z--boson self-energies
(diagrams of Fig.~\ref{fig:oneloopdiagrams})
and neglecting the mixing in the sbottom sector, one obtains for the 
contribution of the $\Stop/\Sbot$ doublet at one--loop order
(only $\Sbot_L$ contributes for $\tsb = 0$) :
\BEA
\De\rho_0^\SU  &=& \frac{3 G_F}{8 \sqrt{2} \pi^2} 
\left[ -\sin ^2 \theta_{\tilde{t}}
\cos^2 \theta_{\tilde{t}} \ F_0( m_{\tilde{t}_1}^2,   m_{\tilde{t}_2}^2) 
\right.
\non \\
&& \left. {} + 
\cos^2 \theta_{\tilde{t}} \ F_0( m_{\tilde{t}_1}^2,   m_{\tilde{b}_L}^2) 
+ \sin^2 \theta_{\tilde{t}} \ F_0( m_{\tilde{t}_2}^2,   m_{\tilde{b}_L}^2) 
\right] ,\\
\mbox{with}\quad
F_0(x, y) &\equiv& x + y - \frac{2 x y}{x - y} \ln\frac{x}{y},
\label{drhooneloop}
\EEA
where $F_0$ has the properties $F_0(x, x) = 0, F_0(x, 0) = x$.
Therefore, the contribution of a squark doublet
becomes in principle very large when the mass splitting between the squarks is
large. This is exactly the same situation as in the case of the SM where
the top/bottom contribution to the $\rho$ parameter at one--loop order
reads~\cite{delrhooneloop} 
\BE
\De\rho_0^{\rm SM}  = \frac{3 G_F}{8 \sqrt{2} \pi^2} F_0(m_t^2, m_b^2).
\EE
For $\mt \gg m_b$ this  leads to the well--known quadratic correction
$\De\rho_0^{\rm SM}= 3 G_F m_t^2 /(8\sqrt{2}\pi^2)$. 

\begin{figure}[ht]
\begin{center}
\mbox{
\psfig{figure=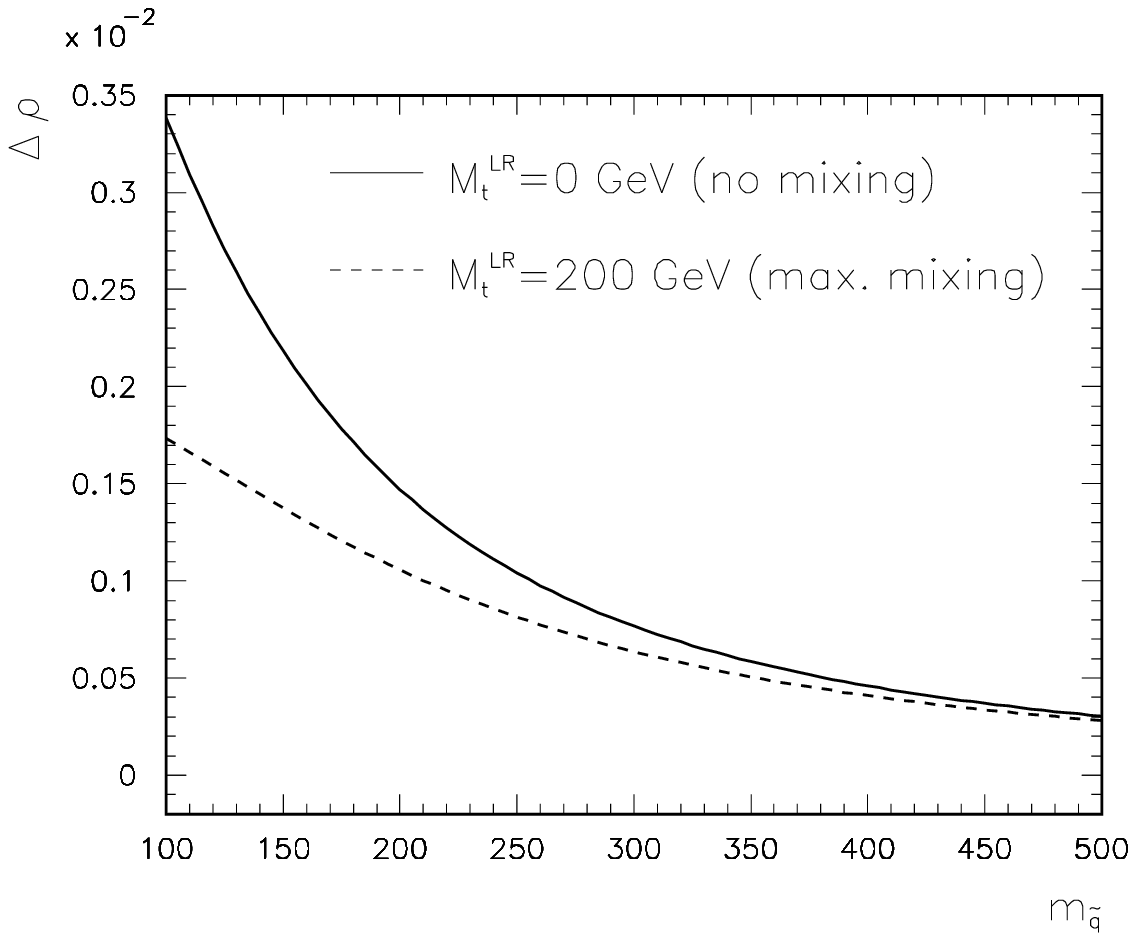,width=10cm,height=7cm,
                                        bbllx=120pt,bblly=270pt,%
                                        bburx=450pt,bbury=540pt}
     }
\parbox{12cm}{
\caption{\label{fig:delrhooneloop}
One--loop contribution of the $(\Stop, \Sbot)$ doublet to
$\De\rho$ as a function of the common squark mass $\msq$ for
$\tst = 0$ and $\tst \sim \pi/4$ (with $\Tb
= 1.6$,  $\mtlr = 0 \mbox{ or } 200  \mbox{ GeV}$).
        }    }
\end{center}
\end{figure}

In Fig.~\ref{fig:delrhooneloop} we display the one--loop correction to
the $\rho$ parameter induced by the $\tilde{t}/\tilde{b}$ isodoublet. 
Due to the SU(2) symmetry and for the sake of simplicity we
concentrated on the 
case $\MstL = \MstR = \MsbL = \MsbR = \msq$.
In this scenario, the scalar top mixing angle is either
very small, $\tst \sim 0$, or almost maximal,
$ \tst \sim  -\pi/4$, in most of the MSSM parameter space.
The contribution $\Delta \rho_0^\SU$ is shown as a function of the common
squark mass $m_{\sq}$ for $\Tb=1.6$ for the two cases 
$\mtlr = 0$ (no mixing) and $\mtlr = 200$ GeV (maximal mixing);
the bottom mass and therefore the mixing in the sbottom sector are
neglected leading to $\msbl = \msbe \simeq \msq$.
Here and in all the numerical calculations in this paper we use 
$m_t=175$~GeV, $M_Z=91.187$~GeV,  $M_W=80.33$~GeV, and $\alpha_s=0.12$. 
The electroweak mixing angle is defined from the ratio of the vector
boson masses: $\sw^2= 1- M_W^2/M_Z^2$. 

\subsection{Two--loop results}\label{subsec:delrho:two}

\begin{figure}[htb]
\begin{center}
\mbox{
\psfig{figure=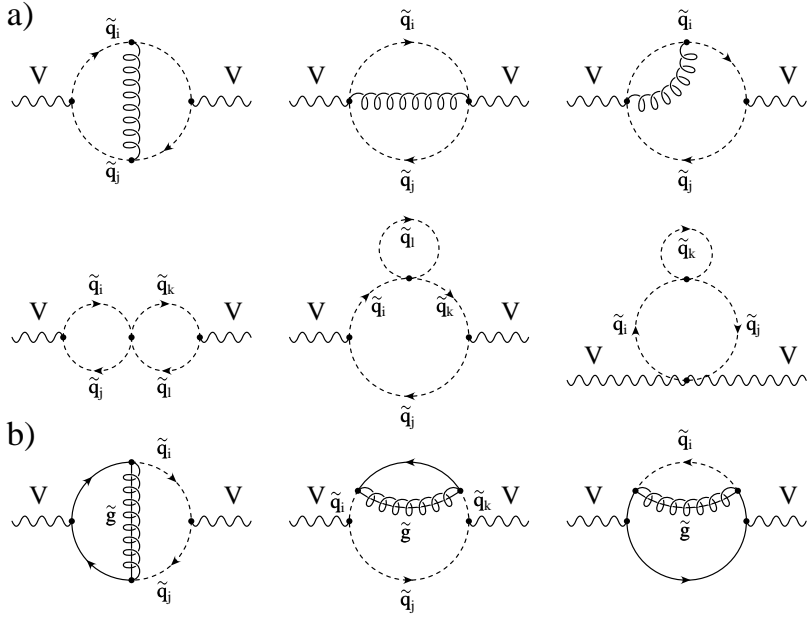,width=10cm,height=8cm,
              bbllx=165pt,bblly=495pt,bburx=450pt,bbury=725pt}
}
\end{center}
\caption[]{\label{fig:twoloopdiagrams}
Typical Feynman diagrams for the contribution of scalar 
quarks and gluinos to the $W/Z$--boson self--energies at the 
two--loop level.}
\end{figure}

The QCD corrections to $\De\rho$ via the squark contributions to the vector
boson self--energies, Fig.~\ref{fig:twoloopdiagrams}, have been
derived recently~\cite{delrholetter,delrholong}. The diagrams  can be
divided into three different classes: the pure scalar diagrams
(Fig.~\ref{fig:twoloopdiagrams}a, 2. line), the gluon--exchange diagrams
(Fig.~\ref{fig:twoloopdiagrams}a, 1. line), and the gluino--exchange diagrams
(Fig.~\ref{fig:twoloopdiagrams}b). These diagrams have to be supplemented
by counterterms for the squark and quark mass renormalization
as well as for the renormalization
of the squark mixing angle. 

The three different
sets of contributions together with their respective counterterms 
are separately gauge--invariant and ultraviolet finite. For the 
gluon--exchange contribution we have only considered the squark loops,
since the 
gluon--exchange in quark loops is just the SM contribution, yielding the result
$\De\rho^{\rm SM}_1 = - \De\rho_0^{\rm SM} \frac{2}{3}
\frac{\al_s}{\pi} (1+\pi^2/3 )$~\cite{delrhotwoloopsm}. 

The pure scalar diagrams give zero contribution: partly they only
contribute to the longitudinal part of the vector boson
self--energies, or they cancel exactly with their corresponding
counterterm diagram. 

\begin{figure}[htb]
\begin{center}
\mbox{
\psfig{figure=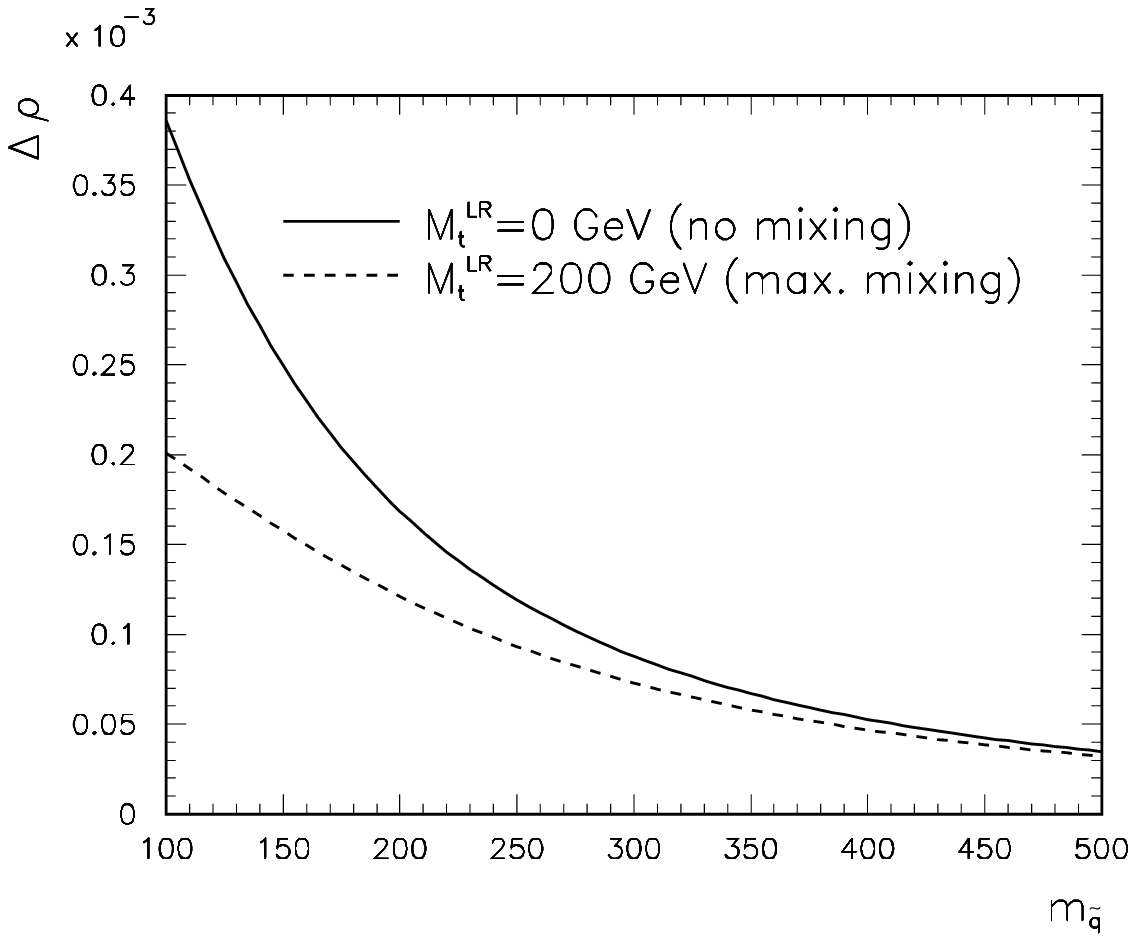,width=10cm,height=7cm,
                                        bbllx=125pt,bblly=270pt,%
                                        bburx=455pt,bbury=540pt}
     }
\parbox{12cm}{
\caption{\label{fig:twoloopgluon}
$\De\rho^\SU_{\rm 1, gluon}$ as a function of $m_{\tilde q}$ for the scenarios
of Fig.~\ref{fig:delrhooneloop}.
        }    }
\end{center}
\end{figure}

The gluon--exchange contribution can be cast into a very simple
formula:
\BEA
\De\rho^\SU_{1, \rm gluon} &=& \frac{G_F \alpha_s}{4 \sqrt{2} \pi^3} \left[ 
- \sin^2\theta_{\tilde{t}} \cos^2\theta_{\tilde{t}}  
F_1\left( \mste^2,  \mstz^2 \right) \right. \nonumber \\ 
&& {} \left. + \cos^2 \theta_{\tilde{t}} 
F_1 \left( \mste^2, \msbl^2 \right)
+\sin^2 \theta_{\tilde{t}}  
F_1 \left( \mstz^2, \msbl^2 \right) \right], 
\label{eq:delrhotwoloopgluon}
\EEA
\BEA
\mbox{with}\quad F_{1}(x,y) &=& x + y - 2\frac{xy}{x-y} \ln\frac{x}{y}
   \left[2 + \frac{x}{y} \ln\frac{x}{y} \right] \non \\
 && +\frac{(x + y) x^2}{(x - y)^2}\ln^2 \frac{x}{y} 
    -2(x - y) {\rm Li}_2(1 - \frac{x}{y})
\EEA
where $F_1$ has the properties $F_1(x, x) = 0, F_1(x, 0) = x(1 + \pi^2/3)$.
The two--loop gluonic SUSY contribution to $\De\rho$ is shown in
Fig.~\ref{fig:twoloopgluon} for the same parameters an in
Fig.~\ref{fig:delrhooneloop}. 

\begin{figure}[htb]
\begin{center}
\mbox{
\psfig{figure=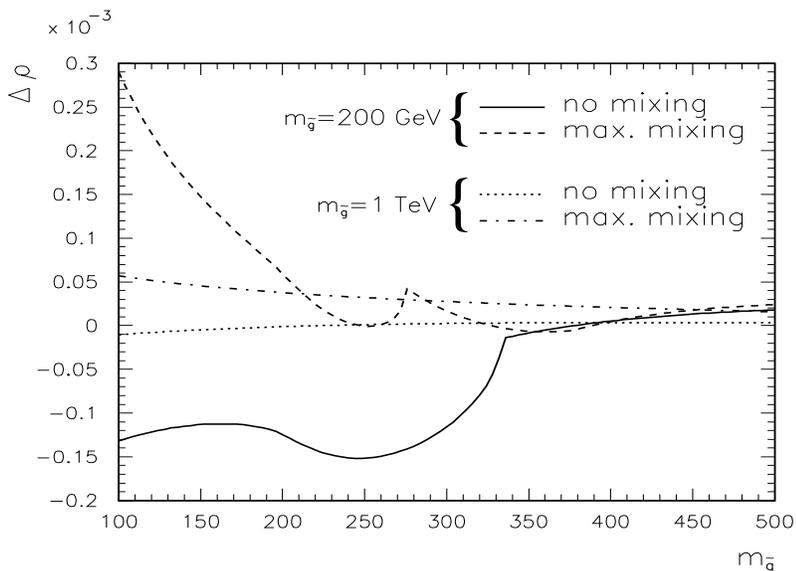,width=10cm,height=7cm,
              bbllx=140pt,bblly=285pt,bburx=450pt,bbury=535pt}}
\parbox{12cm}{
\caption[]{\label{fig:twoloopgluino}
Contribution of the gluino exchange diagrams to
$\De\rho_{1, {\rm gluino}}^{\rm SUSY}$ 
for two values of $\mgl$ in the scenarios of Fig.~\ref{fig:twoloopgluon}.}}
\end{center}
\end{figure}

Like for the gluon--exchange contribution we have also derived a
complete analytic result for the gluino--exchange contribution. This
is, however, very lengthy and we therefore present here a numerical result
in Fig.~\ref{fig:twoloopgluino} for the same parameters as in
Fig.~\ref{fig:twoloopgluon}, 
both for a light and a heavy gluino.
(The complete expression
is available in Fortran and Mathematica format from the author.) 

The complete result has been derived in DREG and DRED. Both methods yield the
same result, since the only diagram which differs in both schemes is the
gluon--exchange in the top self--energy. This, however, contributes to the
SM part only.

To summarize, the gluonic contribution amounts to $\sim 10 - 15\%$ of
the one--loop 
result. Contrary to the SM two--loop corrections, they have the same
sign, resulting in an enhancement in the sensitivity to virtual
squark effects. The gluino--exchange contribution is in general
smaller compared to the gluon result and has different signs. It
competes with the gluon contribution only for $\mgl$ and $\msqi$ close
to their lower experimental bound. There both correction add up to
$\sim 35\%$ for maximal mixing. The gluino contribution decouples
for heavy $\mgl$. We confirmed this by an explicit series expansion in
$\mgl$ starting with an coefficient of ${\cal O}(1/\mgl)$.


\section{QCD corrections to $\De r$}\label{sec:deltar}

$\De r$ fixes the higher order relation between $M_W, M_Z, G_F
\mbox{and } \al$:
\BE
M_W^2 \KL 1 - \frac{M_W^2}{M_Z^2} \KR = 
          \frac{\pi \al}{\sqrt{2} G_F} (1 + \De r),
\EE
where all the radiative corrections are contained in $\De r$. 
At two--loop in leading order the SUSY contribution is given by
\BE
\De r^\SU = \Sigma'_\gamma(0) - \frac{\cw^2}{\sw^2} 
  \KL \frac{\de M_Z^2}{M_Z^2} - \frac{\de M_W^2}{M_W^2} \KR 
   + \frac{\Sigma_W(0) - \de M_W^2}{M_W^2} 
   + 2 \frac{\cw}{\sw}\frac{\Sigma_{\gamma Z}(0)}{M_Z^2}, 
\EE
with $\de M_V^2 = \re \Sigma_V(M_V^2)$ if on--shell renormalization is
used. The SUSY QCD corrections thus only enter via the self--energies.

Since $\De r$ determines the prediction for the mass of the W--boson,
it is crucial to know $\De r$ with high precision.
The complete MSSM one--loop result is available~\cite{deltaroneloop}.
But at least through $\De\rho$~(\ref{eq:delrhoapprox}) also a
non-negligible  two--loop correction is expected.

The diagrams involved here are the same as in the $\De\rho$ case but
in general have non--vanishing external momentum. Therefore they have
a much more complicated structure.
As a first step an analytical result for the gluon--exchange
contribution has been derived. 
Due to the fact that $\mgl \neq 0$, the gluino contribution  can not be 
expressed analytically. Since for large $\mgl$ the gluino decouples,
the gluon--exchange contribution is a good approximation for the case
of a heavy gluino.

The result for a general self--energy $\Sigma_V(k^2)$ at the
two--loop level can be cast into a very compact
form~\cite{deltarlong}, consisting of the scalar one--loop functions
$A_0, B_0$ as well as of the two--loop functions 
$T_{11234'}, T_{1234'}, T_{123'45} \mbox{ and } T_{234'}$. The
$T$--functions can be found in~\cite{lbgroup}. 
After expanding the complete result for $\De r^\SU$ in $\de$, we
found a finite result 
for the gluon--exchange contribution.

\begin{figure}[htb]
\begin{center}
\mbox{
\psfig{figure=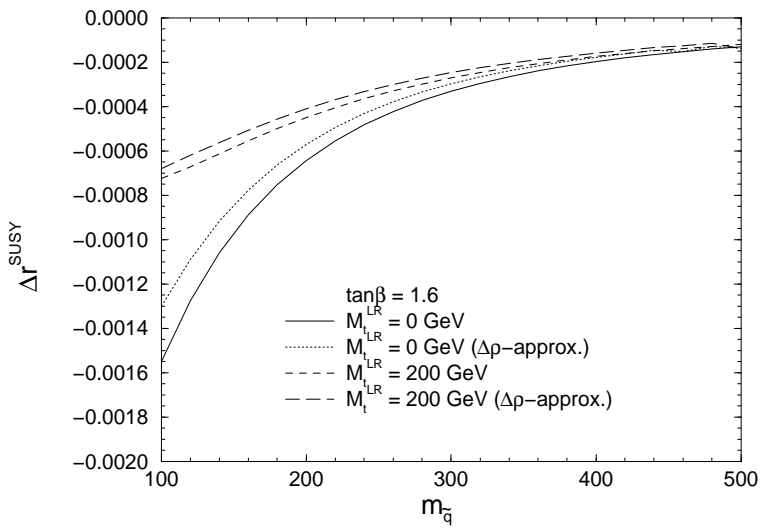,width=10cm,height=7cm,
              bbllx=25pt,bblly=85pt,bburx=550pt,bbury=445pt}}
\parbox{12cm}{
\caption[]{\label{fig:deltartwoloop}
Contribution of the gluon--exchange diagrams to
$\De r^\SU$ and the $\De\rho$--approximation 
given for the scenarios of Fig.~\ref{fig:twoloopgluon}.}}
\end{center}
\end{figure}

The result is presented in Fig.~\ref{fig:deltartwoloop} for the same
parameters as in Fig.~\ref{fig:twoloopgluon}, including also the
$\De\rho$--approximation defined in~(\ref{eq:delrhoapprox}).

The two--loop correction amounts $\sim 10 - 15\%$ of the one--loop
result. The $\De\rho$--approximation reproduces the two--loop
contribution up to $10 - 15\%$.
From this it can be conjectured that also the gluino contribution can be
well approximated with the help of $\De\rho$.
Expressed in terms of the W--mass, the two--loop MSSM contribution
corresponds to a shift in $M_W$ of about $10 - 20$ MeV for light squarks.


\section{Mass of the lightest MSSM Higgs boson: QCD corrections}
\label{sec:higgsmass}

\subsection{Renormalization}\label{subsec:higgsmass:ren}

The MSSM predicts that one of the Higgs bosons is relatively light. At
tree--level $\mh < M_Z$ holds. However, this is changed by radiative
corrections. After taking the one--loop correction into account, 
$\mh > M_Z$ is 
possible, but a new upper bound is set, $\mh < 150$ GeV. At the
two--loop level the upper bound is decreased to about $130$
GeV~\cite{mhiggsupperbound}. Also an RGE calculation exists at the
two--loop level~\cite{mhiggsrge} which shows that $\mh$ is
considerably reduced with respect to the one--loop result.

A diagrammatic calculation has been performed to determine $\mh$ at the
two--loop level for general values of $\Tb$ and $M_A$. In order to
simplify our calculation, we work with 
the unrotated fields $\pe$ and $\pz$ (but we obtain an $\al_{\rm eff}$
at two--loop in the end). We only consider the $t-\Stop$--sector
with no mixing, so we set $\mtlr = 0$. The leading terms are expected
to originate for the $\Phi$ self--energies evaluated for zero external
momentum (which is then similar to the effective potential approach).
We only consider the Yukawa part of the theory,
since due to the high top mass the Yukawa couplings are relatively
large. Thus we set the gauge couplings to zero.

The counterterm for a $\Phi$ self--energy arises only from the Higgs
potential~(\ref{Higgspot}) where we used on--shell renormalization for
the A--boson self--energy. The tadpole renormalization has been chosen
to cancel the tadpole contributions, this leads to
\BEA
\hat{\Sigma}_\Phi(0) &=& \Sigma_\Phi (0) - \de V_\Phi, 
\quad (\hat{\Sigma}: \mbox{renormalized self-energy}) \non \\
\de V_\Phi &=& f_\Phi(\de t_1, \de t_2, \de M_A^2) \\
\mbox{with } && \de t_i = - T_i,\, \de M_A^2 = \Sigma_A(0), \non
\EEA
where $T_i$ denotes the tadpole contribution, $\de t_i$ is the
corresponding counter term.

\begin{figure}[htb]
\begin{center}
\mbox{
\psfig{figure=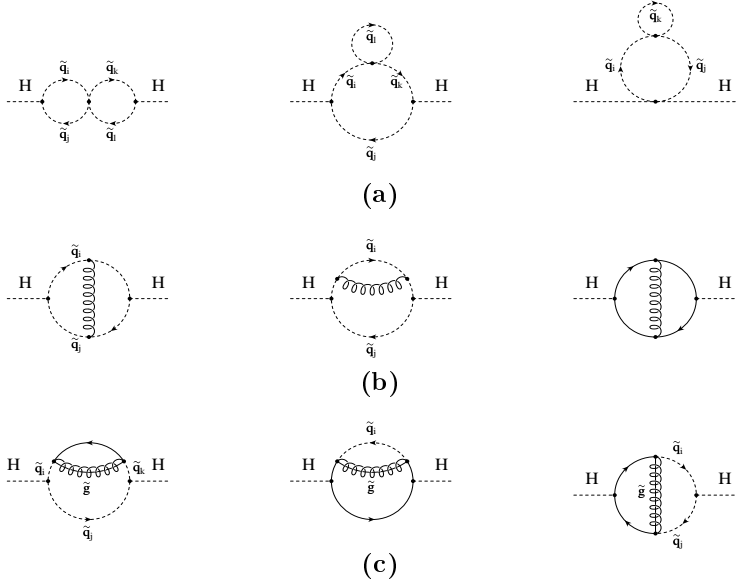,width=10cm,height=8cm,
              bbllx=100pt,bblly=410pt,bburx=500pt,bbury=750pt}
}
\end{center}
\caption[]{\label{fig:higgsdiagrams}
Typical Feynman diagrams for the contribution of quarks and scalar 
quarks with gluons and gluinos to Higgs--boson self--energies at the 
two--loop level.}
\end{figure}

The QCD corrections to $\mh$ via the squark contributions to the $\Phi_i$
self--energies, Fig.~\ref{fig:higgsdiagrams} (we omitted the tadpole
diagrams), are worked out in detail
in~\cite{mhiggslong}. Again the diagrams  can be
divided into three different classes: the pure scalar diagrams
(Fig.~\ref{fig:higgsdiagrams}a), the gluon exchange diagrams
(Fig.~\ref{fig:higgsdiagrams}b), where also the quark loops have to be
taken into account, and finally the gluino--exchange diagrams
(Fig.~\ref{fig:higgsdiagrams}c). These diagrams have to be supplemented
by counterterm insertions for the squark and quark mass renormalization
as well as for a renormalization
of the squark mixing angle. 

Contrary to the $\De\rho$ case the three sets of diagrams do not 
independently give an ultraviolet finite result for the renormalized
$\Phi$ self--energy. In addition it is necessary to make
use of DRED. By explicit calculations both in DREG and in DRED,
we found a finite result only for the DRED. The same observation has
been made also in~\cite{mhiggsupperbound} for the two--loop
calculation of the upper bound for $\mh$.


\subsection{Calculation of $\mh$}\label{subsec:higgsmass:mh}

The mass matrix at two--loop order is given in terms of the tree--level
masses and of the one-- and two--loop
renormalized $\Phi$ self--energies
which can be diagonalized by the angle $\al_{\rm eff}
$\BE
{\cal L}_{2} = 
   \ML m_{\pe}^2 - \hat{\Sigma}_{\pe}^1 - \hat{\Sigma}_{\pe}^2 &
       m_{\pe\pz}^2 - \hat{\Sigma}_{\pe\pz}^1 - \hat{\Sigma}_{\pe\pz}^2\\ 
       m_{\pe\pz}^2 - \hat{\Sigma}_{\pe\pz}^1 -\hat{\Sigma}_{\pe\pz}^2 &
       m_{\pz}^2 - \hat{\Sigma}_{\pz}^1 - \hat{\Sigma}_{\pz}^2 \MR 
   \stackrel{\al_{\rm eff}}{\longrightarrow}
   \ML m_H^2 & 0 \\ 0 & \mh^2 \MR,
\label{eq:mhiggsmatrixtwoloop}
\EE
where $m_{H,h}$ denote the heavy and light Higgs masses at two--loop.

\begin{figure}[htb]
\begin{center}
\mbox{
\psfig{figure=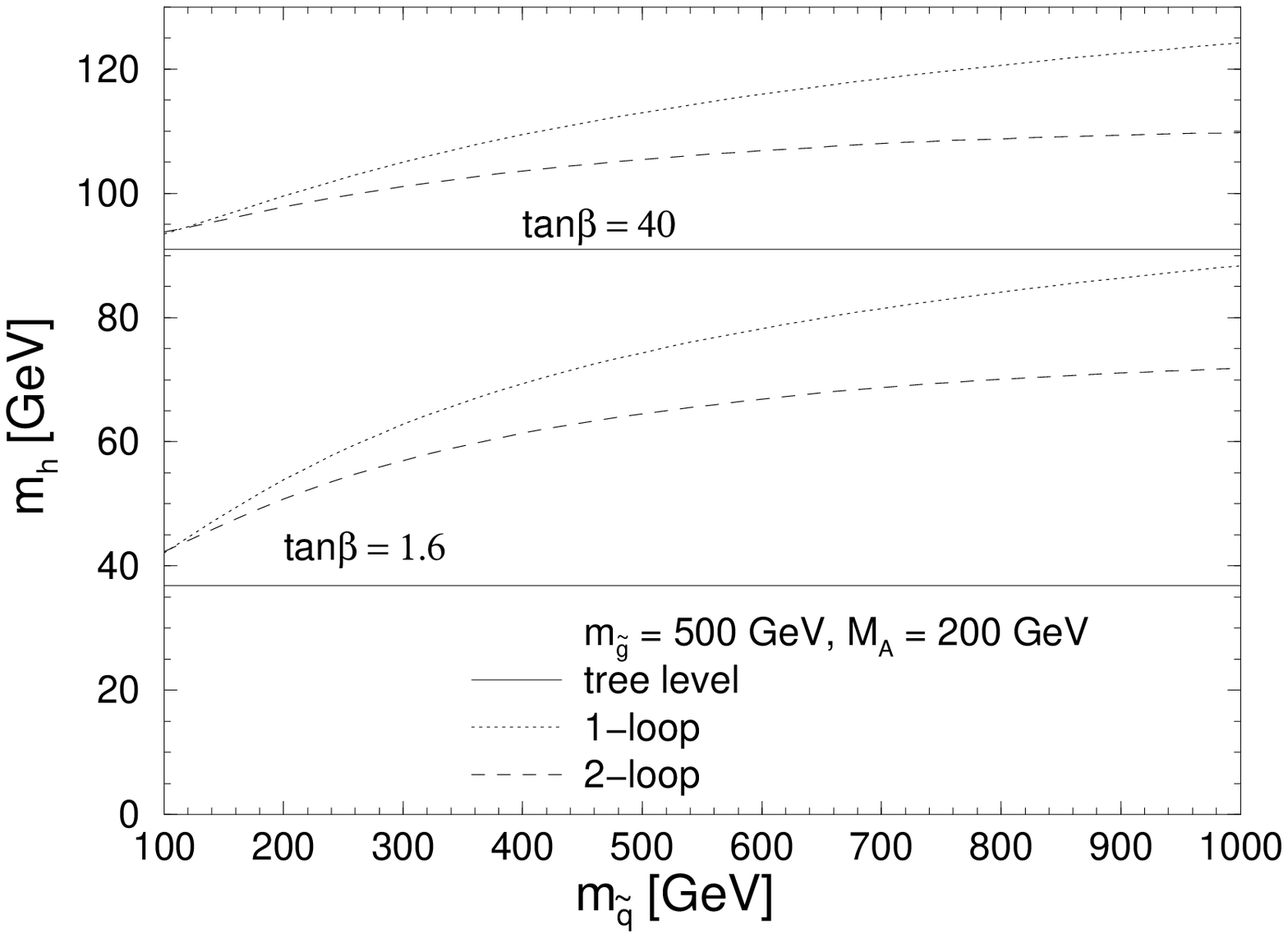,width=10cm,height=7cm,
              bbllx=40pt,bblly=85pt,bburx=530pt,bbury=440pt}
}
\end{center}
\caption[]{\label{fig:higgsmass}
The mass of the lightest MSSM Higgs boson $\mh$ displayed as a
function of $\msq$ for two values of $\Tb$, $\mgl = 500 \mbox{ GeV },
M_A = 200 \mbox{ GeV }$}
\end{figure}

Due to the gluino--exchange contribution the analytical result is very
complicated, it will be given in~\cite{mhiggslong}.
The numerical results are displayed in 
Fig.~\ref{fig:higgsmass} for $\mh$ at the tree--, the one-- and the
two--loop level as a 
function of $\msq$ for the same scenario as in Sec.~\ref{sec:delrho}
but with $\Tb = 1.6$ and $40$, $\mgl = 500$ GeV and $M_A = 200$
GeV ($\mtlr = 0$ GeV.)

The two--loop result lowers the one--loop mass by about $15$ GeV. The
decrease is even somewhat larger for small $\Tb$. This 
strengthens for
the small $\Tb$ scenario, which is favoured by SU(5) GUT scenarios,
the discovery potential of LEP2.


\section*{Acknowledgments}
I want to thank A.~Djouadi, P.~Gambino, C.~J\"unger, W.~Hollik and
G.~Weiglein  with whom I derived the results presented here.


\end{document}